\documentclass{aa}
\usepackage{graphicx,natbib,amsmath,amssymb,fnpara}
\usepackage{txfonts}
\bibpunct[, ]{(}{)}{;}{a}{}{,}
\def\OM{\Omega_{\rm M}}
\def\OX{\Omega_{\rm X}}
\def\OL{\Omega_{\Lambda}}
\def\Ep{E_{\rm peak}}
\def\Eg{E_{\gamma}}
\def\OX{\Omega_{\rm X}}
\def\gtrsim{\mathrel{\hbox{\rlap{\hbox{\lower4pt\hbox{$\sim$}}}\hbox{$>$}}}}

\begin{document}

\title{On the future of Gamma-Ray Burst Cosmology}

\titlerunning{GRB cosmology}

\author{E.~M\"ortsell\inst{1}
        \and J.~Sollerman\inst{1} }

\offprints{edvard@astro.su.se}

\institute{Department of Astronomy, Stockholm University,
AlbaNova, 106 91 Stockholm, Sweden}

\date{Received / Accepted}

\abstract{
With the understanding that the enigmatic Gamma-Ray Burts (GRBs) are
beamed explosions, and with the recently discovered
``Ghirlanda-relation'', the dream of using GRBs as cosmological
yardsticks may have come a few steps closer to reality. Assuming the
Ghirlanda-relation is real, we have investigated possible constraints
on cosmological parameters using a simulated future sample of a large
number of GRBs inspired by the ongoing {\tt SWIFT} mission. Comparing
with constraints from a future sample of Type Ia supernovae, we find
that GRBs are not efficient in constraining the amount of dark energy
or its equation of state. The main reason for this is that very few
bursts are available at low redshifts.
}

\maketitle
%
\section{Introduction\label{introduction}}

The usage of thermonuclear Type Ia supernovae (SNe Ia) has
revolutionized cosmology. These intrinsically bright explosions are
almost standard candles in optical light. With a simple light-curve
correction they can be standardized to high enough precision to probe
in detail the energy content of the universe. This led to the
discovery that dark energy dominates the presently accelerating
universe \citep[e.g.,][]{riess98,perlmutter99}. Today, SN Ia data is
approaching the quantity and quality where it is possible to constrain
not only the amount of dark energy but also dark energy properties
\citep[e.g.,][]{hannestad02,hannestad04}.

Gamma-Ray Bursts are even more powerful explosions. They have recently
been firmly linked to energetic core-collapse supernovae
\citep{hjorth03,matheson03}. Their isotropic energy appears to
outpower thermonuclear supernovae, which allows studies at even higher
redshifts. Also, in contrast to the case of SNe Ia where the rate is
unknown at $z\gtrsim 1.5$ \citep{dahlen04}, we know that GRBs exist at
high redshifts \citep[e.g.,][]{andersen00}. Moreover, the released
burst of gamma-rays can penetrate the dust that obscures our view of
the distant universe in optical light. For current constraints on dust
attenuation of SNe Ia, see \cite{ostman05}. The potential of GRBs as
probes for cosmological investigations thus appears to be very good.

The realization that the total gamma-ray energy of a GRB, when
corrected for the effects of beaming, spans a reasonably narrow range
of energies \citep{frail01} arose hope for GRB cosmology
\citep[e.g.,][]{schaefer03}. More recently, the discovered tight
relation between the rest-frame peak energy $\Ep$ of the GRB and the
rest-frame, beaming-corrected gamma-ray energy release $\Eg$, the
so-called ``Ghirlanda-relation'' \citep{ghirlanda04a}, has renewed the
hope for GRB cosmology. This relation allows an empirical correction
to the determined luminosity distances for each GRB, in much the same
way as light-curve shape corrections are applied to SNe Ia, with a
scatter along the GRB Hubble diagram of $\sim0.5$ mag
\citep{ghirlanda04b}. This has caused a fierce activity of research
in the area of GRB cosmology
\citep[e.g.,][]{ghirlanda04b,dai04,firmani05,xuetal05,xu05}.

There is, however, much to be done before GRB cosmology can be
established. The reality of the Ghirlanda relation is still under
discussion \citep{bandpreece05,friedmanbloom05,ghirlanda05}.
Moreover, very different results on GRB cosmology are obtained by the
different authors, based on the current sample of GRBs.
\cite{friedmanbloom05} thoroughly examined pro and cons
of GRB cosmology with the presently limited sample of well studied
GRBs, and show that the obtained results are crucially dependent on
the choice of (poorly known) input parameters. The way to select which
GRBs to include also influence the results. \cite{friedmanbloom05} are
therefore rightfully cautious concerning claims of the utility of GRBs
for cosmology. A larger dataset is clearly needed to establish this
issue.

The recently launched {\tt SWIFT} satellite \citep{gehrels04} will
find hundreds of GRBs. In this paper we simulate the potential effect
of such a large number of GRBs for the use of determining cosmological
parameters, assuming that the Ghirlanda relation holds.

In Sect.~\ref{method} we present the method used for deriving
cosmological parameters from GRBs.  We also discuss various
assumptions for the constructed input samples that we use for the
simulations. The results are discussed in Sect.~\ref{discussion}.

%
\section{Simulations\label{method}}

\subsection{The method\label{methodII}}

In the following, we have basically followed the formalism laid out
in, e.g., \cite{friedmanbloom05}. GRBs are used in much the same way
as SN Ia standard candles in constraining cosmological parameters,
i.e, by comparing observed luminosity distances to theoretical
predictions. However, since the process of standardizing the GRB
candles is cosmology dependent, the Ghirlanda relation needs to be
recalibrated for each cosmology. Effectively, this amounts to
refitting for each cosmology the $\Ep-\Eg$ power-law

\begin{equation}
\Ep = \kappa\left(\frac{\Eg}{E_{0}}\right)^{\eta} .
\end{equation}

Note that $E_0$ is an arbitrary constant and that by putting
$E_0\propto h^{-3/2}$, the best-fit $\kappa$ and $\eta$ will be
independent of the value of the Hubble parameter. Thus, marginalizing
over $\kappa$ and $\eta$ is similar to marginalizing over ${\mathcal
M}$ for SNe Ia \citep[e.g.,][]{hannestad02}.

There have been several suggestions on how to include the information
from the $\Ep-\Eg$ fit in the cosmology fit. \cite{dai04} basically
ignored this complication.  This was quickly noted and remedied
by \cite{ghirlanda04b}, who refitted the relation for each cosmology.
\cite{ghirlanda04b}  and \cite{xuetal05} have also
used a simple, but rather unrealistic, approach in assuming that
$\kappa$ and $\eta$ can been exactly determined using a sample of low
redshift GRBs, or by theoretical considerations, and does not need to
be recalibrated. \cite{friedmanbloom05} also refits the $\Ep-\Eg$
power-law and obtains a new set of $\kappa$ and $\eta$ with
corresponding errors for each cosmology. These errors are then
propagated to the error in the derived luminosity distance. This
approach has the drawback of giving smaller $\chi2$-values for
cosmologies where the $\Ep-\Eg$ relation is badly fit since the
luminosity distance errors are larger. \cite{xu05} discuss a method
where cosmologies with a good fit of the $\Ep-\Eg$ relation are
favored (their Method III). This method is similar to adding the
$\chi2$-values from the cosmology fit and the power-law fit for each
cosmology.

In this paper, we treat $\kappa$ and $\eta$ as unknown parameters that
should be marginalized over. This means that we do not use the
$\kappa$ and $\eta$ that gives the best fit to the $\Ep-\Eg$ relation
for each cosmology, but instead use the $\kappa$ and $\eta$ that
minimizes the cosmology $\chi2$-value, in analogy with marginalizing
over ${\mathcal M}$ for SNe Ia. We have noted that our constraints on
the cosmological parameters does not depend sensitively on the
specific method used, as long as $\kappa$ and $\eta$ are not assumed
to be fixed.

\subsection{The GRB sample\label{sample}}

The aim of this paper is to investigate the future potential of GRB
cosmology by simulating a larger sample of GRBs. For supernova
cosmology, many such investigations have been performed, in particular
in conjunction with the planned {\tt SNAP} mission
\citep[e.g.,][]{goliath01,wang04}.

To simulate the effect of a larger future sample we have chosen 200
GRBs with essentially similar properties as the current sample. This
number of GRBs is in line with the expectations on the {\tt SWIFT}
satellite, which is predicted to find about 100 bursts every year for
a life time of 2-8 years \citep{gehrels04}.

\subsubsection{The redshift distribution}

An important property of the future sample is the redshift
distribution of the GRBs. It has been suggested that GRBs and their
afterglows can be detected up to very high redshifts
\citep{lamb02}. To investigate the dependence of the cosmological
predictions on the distribution of the objects we have used two
published predictions for the redshift distribution.

First, the calculations performed by \cite{brommloeb02} suggest that
the {\tt SWIFT} satellite will be able to detect GRBs up to redshifts
of $z \gtrsim 20$. They find that 25$\%$ of the detected bursts will
have $z>5$. We have adopted their calculated redshift distribution for
{\tt SWIFT} as input for our simulations.

\begin{figure}[t]
\includegraphics[width=65mm,clip]{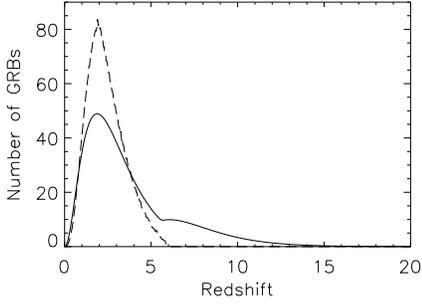}
\caption{The redshift distributions used for our simulated GRB sample.
The dashed line is the predicted redshift distribution available to
{\tt SWIFT} according to \cite{gorosabel04}, while the solid line
shows the distribution of GRBs according to the calculations by
\cite{brommloeb02}.}
\label{f:distribution}
\end{figure}

Secondly, a more conservative estimate is performed by
\cite{gorosabel04}. They argue that relatively few high-redshift
GRBs will be found by {\tt SWIFT}, and calculating the expected
redshift distribution from their Fig.~3 we distribute the simulated
GRBs up to $z\sim 6$.  These two distributions, as shown in
Fig.~\ref{f:distribution}, can thus be taken to represent two extremes
for a future GRB sample.

We note that systematic effects such as
gravitational lensing are potentially greater at high redshifts
\citep{bergstrom01,amanullah03}. In this study,
we assume these errors to be negligible. Note also that the
probability for multiple lensing increases with redshift and that
multiple imaged GRBs therefore are potentially useful for constraining
the Hubble parameter and galaxy halo properties
\citep{goobar02b,mortsell05}.

\subsubsection{The properties of the sample}

We have taken the current sample of 19 GRBs from the list provided by
www.comsicbooms.net, and distributed the observables and their
associated errors in the simulated sample in the same way as for the
observed sample.

The fractional errors are, in order of decreasing importance for the
total magnitude error, the peak energy $\sigma_{\Ep}/\Ep=0.176$, the
{jet break time} $\sigma_t/t=0.18$, the {cirumburst} medium density
$\sigma_n/n=0.5$, the fluence $\sigma_S/S=0.1$ and the k-correction
$\sigma_k/k=0.056$. The contributions to the magnitude error are
$\sigma_m^{\Ep}\sim 0.4$, $\sigma_m^{t}\sim \sigma_m^{n}\sim 0.2$,
$\sigma_m^{S}\sim 0.1$ and $\sigma_m^{k}\sim 0.06$. The total
magnitude error is $\sigma_m\sim 0.5$.

We have chosen to assume that the uncertainties in the simulated
future sample will be the same as in the current sample of GRBs. This
is similar to the approach by \cite{xuetal05}. We also assume that
even the most distant bursts will be followed in enough detail not to
deteriorate the sample. In the gamma-ray regime, it is clear that the
intrinsic scatter in parameters is already larger than expected from
the redshift distribution alone.

For a true {\tt SWIFT} sample we should distribute the $\Ep$ in a
smaller range, since the BAT instrument is sensitive only in the
relatively narrow 15-150~keV range.  We have, for the sake of
simplicity, ignored this matter, but note that it has been used as an
argument to continue the efforts for {\tt HETE\,II} and {\tt Integral}
\citep{friedmanbloom05,gorosabel04}.

\cite{lamb02} argued that even for the most distant GRBs, follow-up
observations will be possible. Redshift determinations will indeed be
feasible up to $z\sim 10$ with instruments such as X-shooter
\citep{dodorico04}. While optical follow-up of very distant bursts
may not be good enough to probe the jet break in detail, also X-ray
and near-IR facilities are available for this. \cite{gorosabel04}
estimate that a near-IR afterglow can be followed up to $z=9$ even
with modest exposure times on a 10-m class telescope.  The very late
occurrences of jet-breaks will of course be difficult to detect for
the dimmest targets, which may bias the sample.

The narrow energy range for SWIFT means that 200 GRBs with
measurements of all relevant observables is probably a too optimsitic
assumption for this mission. This is also supported by the fact that
the SWIFT bursts detected so far appears to be faint and has therefore
not been successfully characterised \citep{berger05}.  The sample we
use for our simulations should therefore be regarded as a rather
optimistic guess that has been influenced by the SWIFT mission, but
may have to await future missions to be accomplished. This optimsitic
assumption will only strengthen our conclusions given below.

\subsection{The SN sample\label{SNsample}}

To compare the constraints on the cosmological parameters obtained
from our GRB simulation we have also simulated a future set of SN Ia
data using the publicly available SNOC package \citep{goobar02c}. Just
as the constructed GRB sample was motivated by the {\tt SWIFT}
mission, which is already ongoing, we base the constructed SN Ia
sample on the available Gold sample \citep{riess04} as well as two
ongoing supernova surveys, the ESSENCE project and the Supernova
Factory.

The ESSENCE project \citep[e.g.,][]{matheson05} is an ongoing survey
aimed to measure 200 SNe Ia in the redshift domain $z=[0.2-0.8]$. The
goal is to derive tight constraints on the equation of state, $w$, of
the dark energy.  In constructing the sample for our simulations we
have used the redshift distribution of the hitherto discovered 109 SNe
Ia, but increased the number of SN Ia to 200, as planned for ESSENCE
\citep{miknaitis04}.
We have adopted an intrinsic photometric error for an individual
supernova of 0.14 mag. Binning 200 SNe into $\Delta m=0.1$ redshift
bins gives a statistical uncertainty
\citep{garnavich02,miknaitis05}
close to the systematic floor expected from e.g., uncertainties in the
k-corrections.

The Supernova factory is an ongoing effort to measure 300 nearby,
$z=[0.03-0.08]$, SNe Ia \citep[e.g.,][]{aldering02}. We have assumed
an uniform redshift distribution in this range, and again an
individual error of 0.14 mag per supernova. Note that our approach
only includes statistical errors.

\section{Results and discussion}\label{discussion}

\begin{figure}[t]
\setlength{\unitlength}{1mm}
\begin{picture}(120,120)(0,0)
\put (0,58) {\includegraphics[width=57mm,bb=54 360 479 785,clip]{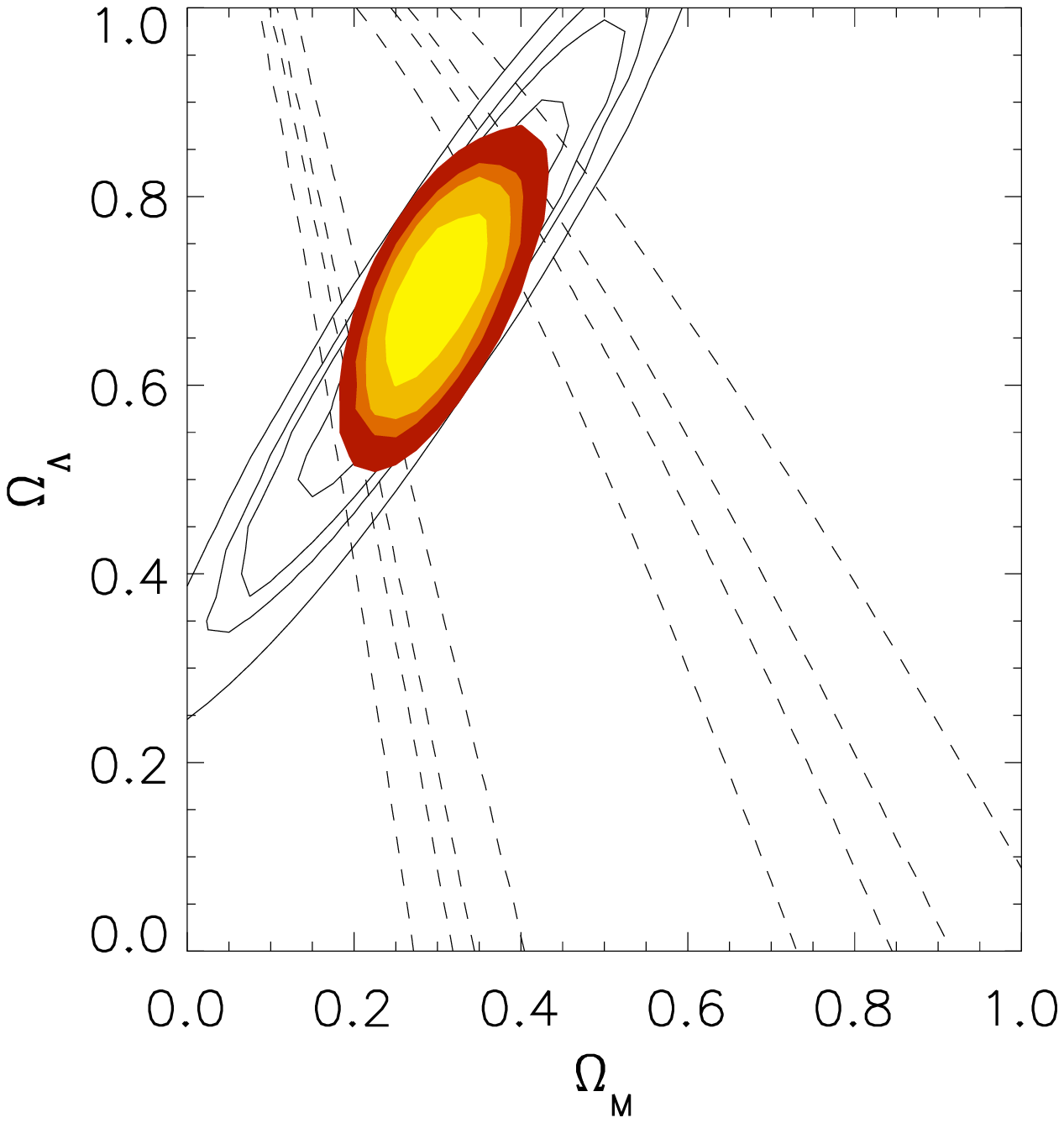}}
\put (58,58){\includegraphics[width=57mm,bb=54 360 479 785,clip]{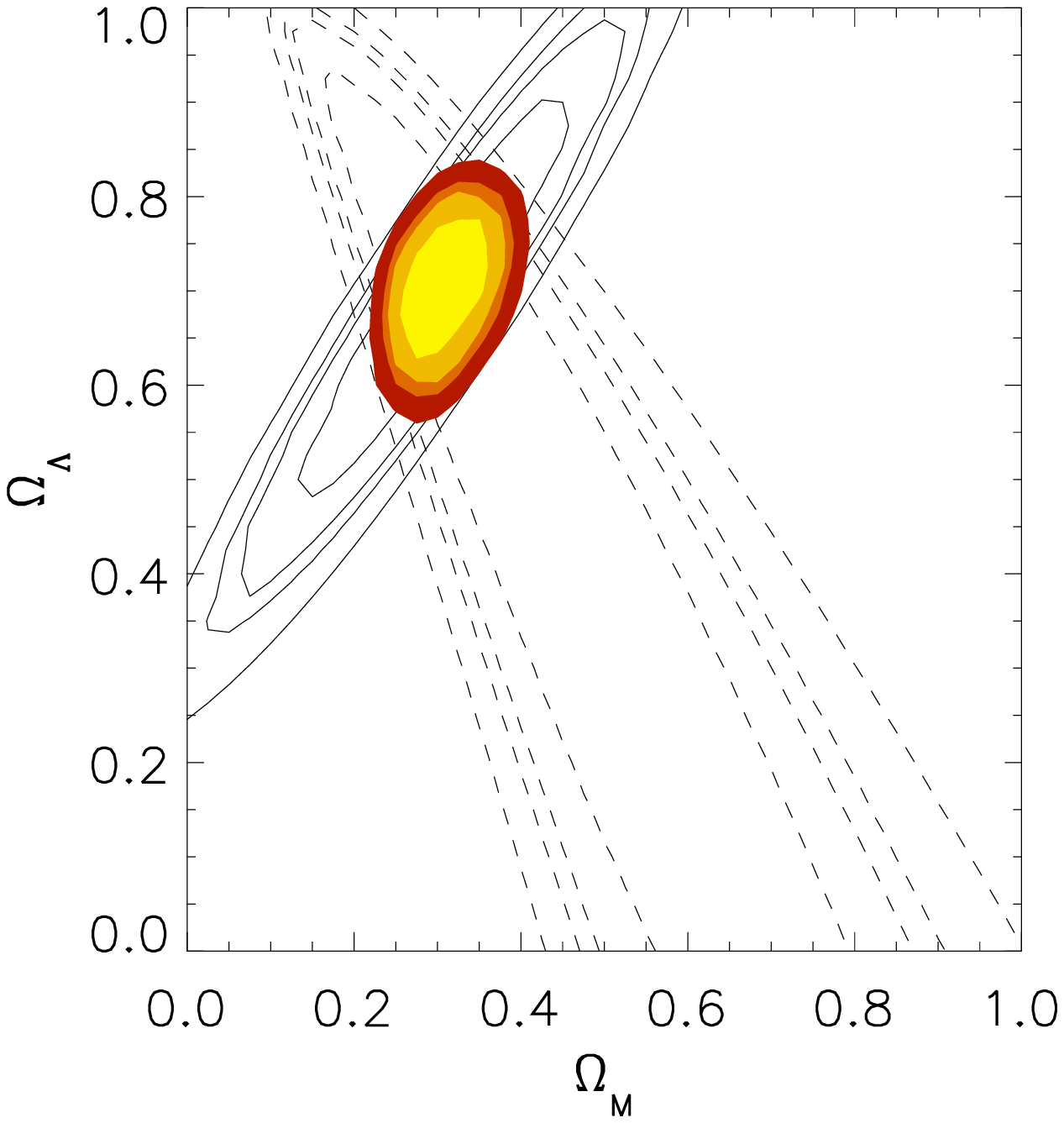}}
\put (0,0) {\includegraphics[width=57mm,bb=54 360 479 785,clip]{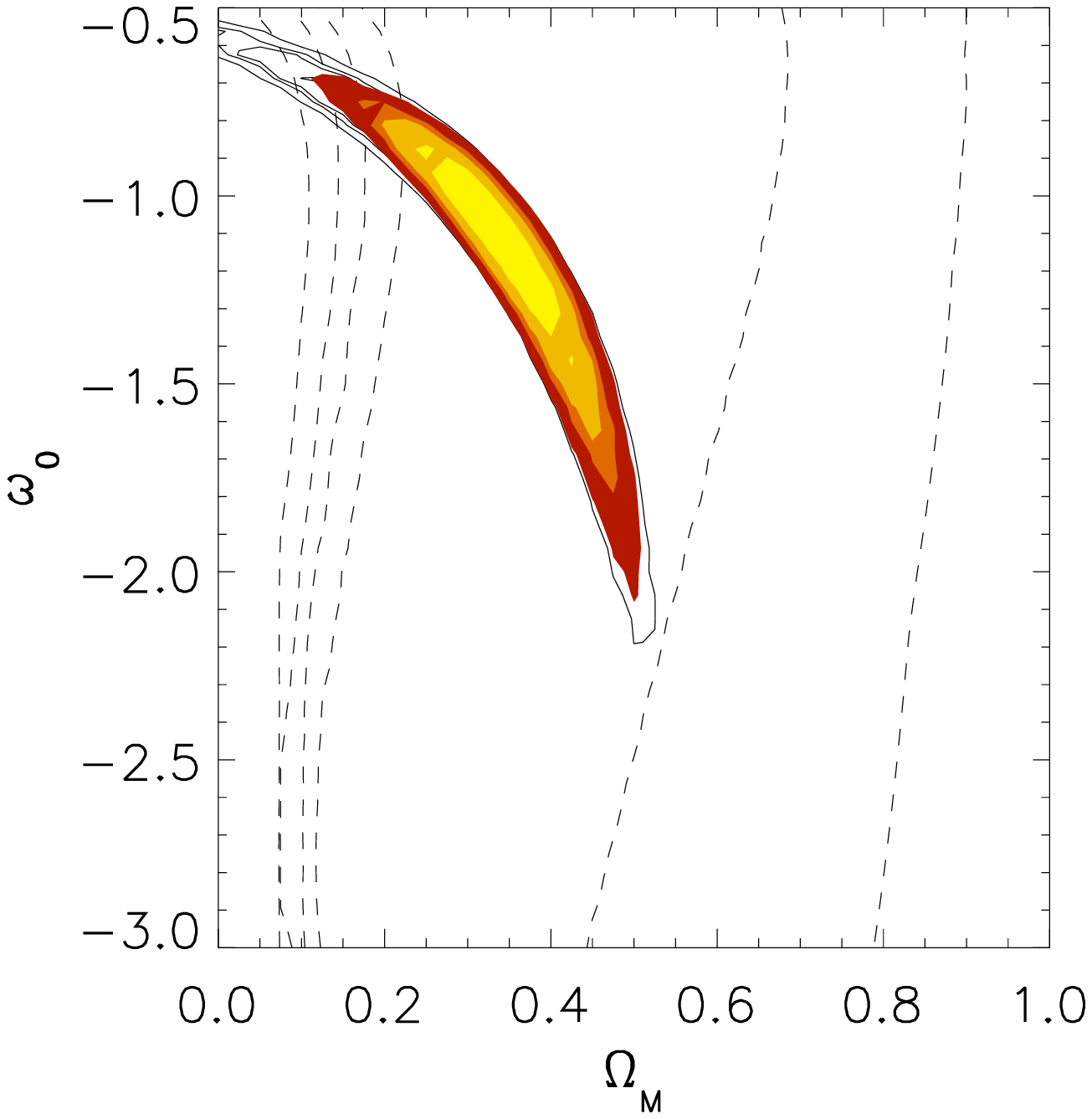}}
\put (58,0) {\includegraphics[width=57mm,bb=54 360 479 785,clip]{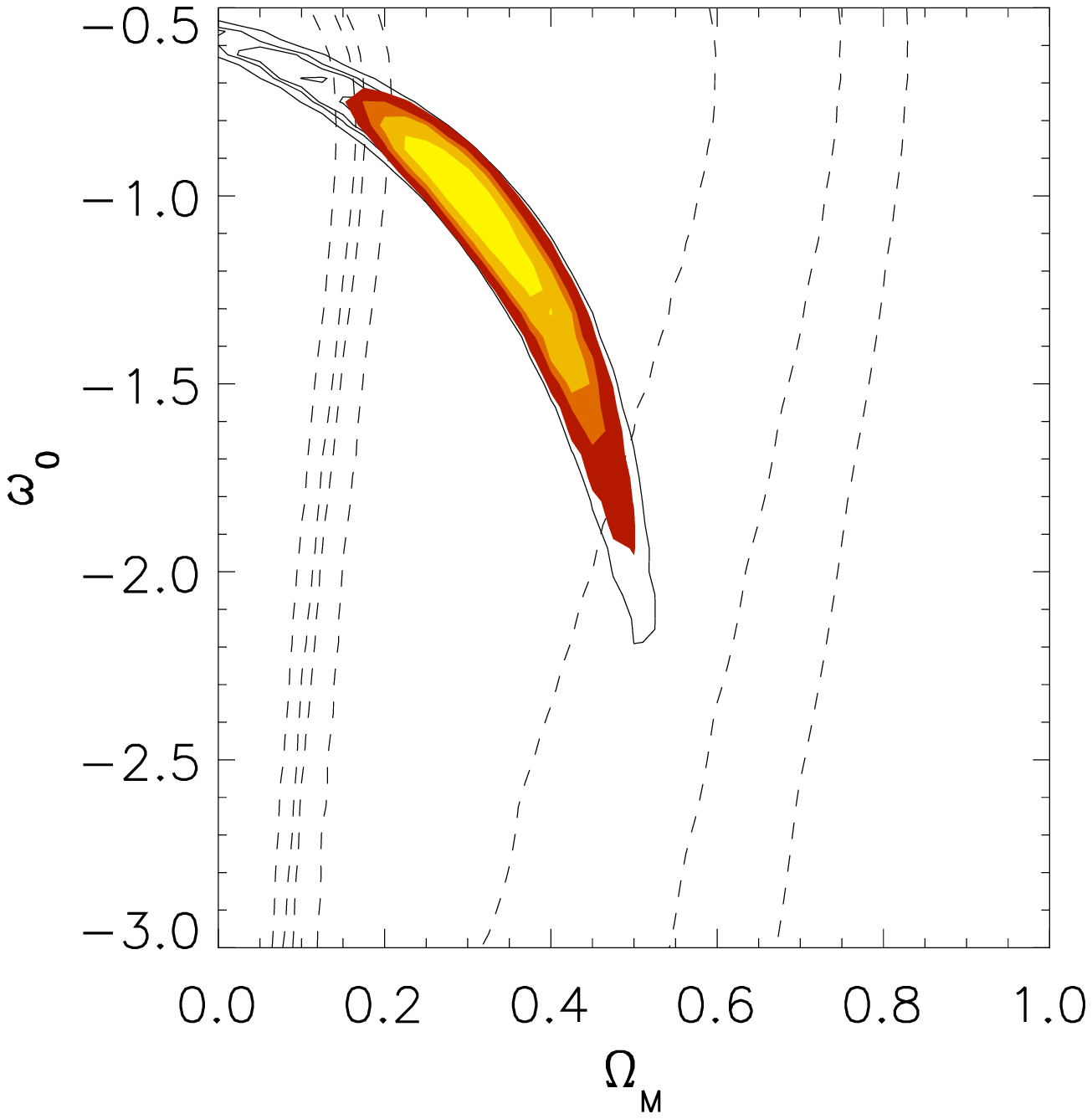}}
\end{picture}
\caption{
{\it Upper left:} $\OL$ versus $\OM$ for the GRB sample based on the
\cite{gorosabel04} redshift distribution (dashed lines) and the SN
sample (solid lines). Combined results are showed in yellow (filled
contours). {\it Upper right:} $\OL$ versus $\OM$ for the GRB sample
based on the \cite{brommloeb02} redshift distribution.  {\it Lower
left:} $w_0$ versus $\OM$ assuming a flat universe for the GRB sample
based on the \cite{gorosabel04} redshift distribution (dashed lines)
and the SN sample (solid lines). Combined results are showed in yellow
(filled contours).  {\it Lower right:} $w_0$ versus $\OM$ for the GRB
sample based on the
\cite{brommloeb02} redshift distribution. }
\label{f:all}
\end{figure}

In Fig.~\ref{f:all} we provide the results of our simulations.  All
simulations have been made assuming a flat universe ($\OM=0.3$,
$\OL=0.7$) dominated by a cosmological constant ($w=-1$).

In Fig.~\ref{f:all} we show the probability contours for the 200 GRBs
(dashed lines) and the combined (Gold sample+ESSENCE+SN factory) SN Ia
sample (solid lines). Combined results are showed using yellow (filled
contours). The four contours correspond to $68.3\,\%$, $90\,\%$,
$95\,\%$ and $99\,\%$ confidence, respectively.

From the top row in Fig.~\ref{f:all}, it can be seen that the GRB
sample (dashed lines) in itself is not very efficient in constraining
$\OL$. This is true for both adopted redshift distributions. The
interesting aspect is that the constraints are rather complementary to
the SN Ia constraints, which makes the combination of these two
datasets relatively fruitful. That the GRB cosmology is predominantly
sensitive to $\OM$, and rather insensitive to the value of the
cosmological constant, reflects the higher redshift distribution of
this sample. At these epochs, the dark energy contribution to the
energy density was still relatively unimportant and we lack a set of
low-$z$ GRBs that would provide the necessary leverage in the Hubble
diagram to constrain the dark energy.

Clearly, a good prior on $\OM$ from large scale structure (LSS)
surveys, such as 2dF
\citep{percival01} or SDSS \citep{tegmark04} are more helpful in constraining
the SN Ia contours, but we find it noteworthy that the same thing can
be achieved using only standard candle techniques.

The lower row in Fig.~\ref{f:all} shows constraints derived for a
constant equation of state parameter, $w=w_0$, for the dark
energy. When fitting $w_0$, we have assumed a flat universe, i.e.,
$\OM + \OX = 1$. The solid contours correspond to the SN constraints
and the dashed contours to GRB constraints. From this exercise it can
be seen that GRB cosmology can not be expected to give very useful
constraints on the equation of state parameter. It has been suggested
\citep[e.g.,][]{friedmanbloom05} that the GRBs could be useful to
constrain any potential evolution of $w$. We believe that, given the
large uncertainties in even constraining a constant value for the
equation of state parameter, such efforts will be in vain. It is
interesting to note that combined with a good prior on $\OM$ from,
e.g., LSS surveys, SN Ia data alone will provide very powerful
constraints on a constant equation of state parameter within a few
years.

Our main conclusion is thus that GRBs are not likely to contribute
significantly to constraints on the cosmological parameters in the
near future. This is similar to the wordings in
\cite{friedmanbloom05} and the also to the results of \cite{xuetal05}
when not assuming an artificially fixed $\Ep-\Eg$ relation but at odds
with other investigations \citep[e.g.,][]{ghisellini05,lazzati05}.
The main reasons for the limited use of GRBs, apart from the huge
inherent uncertainties in trying to standardize these candles
\citep[e.g.,][]{friedmanbloom05}, is their redshift
distribution. Probing $\OX$ and its equation of state benefits from a
wide redshift distribution, with good coverage at the epochs where
$\OX$ dominates the energy density and a set of low-$z$ events to
constrain the normalization of the GRB luminosity. By artificially
including a large number of low-$z$ bursts into the sample, we were
indeed able to considerably shrink the confidence regions for the
cosmological parameters. The point is that the calculations used for
the future GRB redshift distributions used in this work predict that
very few of these rare bursts will be available within the limited
volume spanned by these low redshifts.\\

{\em Acknowledgements.}  We want to thank Javier Gorosabel and Volker
Bromm for the kindness of sending us their calculations of redshift
distributions.  We are also very thankful to the Cosmicbooms.net for
collecting the properties of GRBs and making them public. We further
thank Jens Hjorth, Johan Fynbo, Felix Ryde, Claes Fransson,
Claes-Ingvar Bj\"ornsson and others in Stockholm for interesting
discussions and the anonymous referee for helpful comments.

{}

\clearpage

\end{document}